\documentclass[aps,superscriptaddress,twocolumn,twoside,floatfix,pra,a4paper]{revtex4-2} %nofootinbib

\usepackage{times}
\usepackage{epsfig}
\usepackage{amsfonts}
\usepackage{amsmath}
\usepackage{amssymb,amsthm}
\usepackage{xcolor,colortbl}
\usepackage{multirow}
\usepackage{braket}
\usepackage{latexsym}
\usepackage{amsfonts}
\usepackage{mathrsfs}
\usepackage{natbib}
\usepackage{verbatim}
\usepackage{gensymb}
\usepackage{caption}
\usepackage{caption}
\usepackage{subcaption}
\usepackage{ragged2e}

\usepackage{soul}
\DeclareCaptionJustification{justified}{\justifying}
\usepackage{blkarray}
 \usepackage{graphicx}
\newtheorem{theorem}{Theorem}

\newtheorem{definition}{Definition}
\newtheorem{proposition}{Proposition}

\usepackage[colorlinks=true,linkcolor=blue,citecolor=magenta,urlcolor=blue]{hyperref}
\allowdisplaybreaks

\newtheorem{Lemma}{Lemma}

\begin{document}

\title{One-time Pad Encryption Model for Non-local Correlations}

\author{Govind Lal Sidhardh}
\affiliation{Department of Physics, Duke University, Durham, North Carolina, USA 27708}

\author{Manik Banik}
\affiliation{Department of Physics of Complex Systems, S.N. Bose National Center for Basic Sciences, Block JD, Sector III, Salt Lake, Kolkata 700106, India.}

\begin{abstract}
We present a cryptographic-inspired framework for modeling Bell nonlocal correlations. Drawing inspiration from the renowned De Broglie-Bohm theory, we conceptualize nonlocal boxes as realistic systems featuring instantaneous signaling at the hidden variable level. By introducing randomness into the distribution of the hidden variable the superluminal signaling model is made compatible with the operational no-signalling condition. As our design mimics the famous symmetric key encryption system called {\it One-time Pad} (OTP), we call this the OTP model for nonlocal boxes. We illustrate the efficacy of this model through various esoteric examples related to the non-classical nature of nonlocal boxes. In particular, the breakdown of communication complexity using nonlocal boxes can be better understood in this framework. Additionally, we delve into the Van Dam protocol, revealing its connection to homomorphic encryption studied in cryptography. Exploring potential avenues for encapsulating quantum-realizable nonlocal correlations within our framework, we highlight that the Information Causality principle imposes additional constraints at the hidden variable level. Present work thus orchestrates the results in classical cryptography to improve our understanding of nonlocal correlations and welcomes further research to this connection.    
\end{abstract}

\maketitle
\section{Introduction} 
Violation of Bell inequality drastically revolutionizes our world-view \cite{Bell1964} (see also \cite{Bell1966,Mermin1993,Brunner2014}). An input-output correlation with spatially separated input settings cannot have a local hidden variable model whenever it violates a Bell inequality, provided the input settings are independent of the hidden variable. This nonlocal behavior creates an apparent tension with the well-regarded no-signaling (NS) principle \cite{Maudlin1994}. However, by introducing randomness into the hidden variable model, a peaceful coexistence between these two concepts becomes possible. A celebrated example is De Broglie-Bohm theory \cite{Broglie1928,Bohm1952(1),Bohm1952(2)}, which accommodates both nonlocality and NS by adhering randomness in the distribution of nonlocal variable \cite{Valentini1991}. 

While Bell's theorem offers a fundamentally new outlook of the physical world, a better appreciation of the Bell inequality emerged when it was realized that the non-local quantum correlations inherent in entangled particles could be utilized as cryptographic keys \cite{Ekert1991}. Subsequently, this intuition underwent a more detailed analysis leading to a comprehensive establishment on a solid foundation \cite{Barrett2005a,Acin2006,Acin2007,Vazirani2014}. A number of experiments involving entangled quantum systems establish our world to be non-local \cite{Clauser1969,Freedman1972,Aspect1976,Aspect1981,Aspect1982,Aspect1982(1),Zukowski1993,Pan1998,Weihs1998,Bouwmeester1999} (see also \cite{Wiki} for the list of other important Bell tests). Later, it is realized that nonlocality is not a feature specific to quantum theory only. Popescu \& Rohrlich come up with a hypothetical NS correlation \cite{Popescu1994}, now familiar as PR box, which turns out to be more non-local than quantum theory \cite{Cirelson1980}. Such a post-quantum non-local correlation has preposterous consequences. For instance, the existence of the PR box in a world leads to the collapse of communication complexity, as any non-local binary function could be evaluated with a constant amount of communication when assisted with such  correlations \cite{vanDam2005,Brassard2005,Buhrman2010}. In addition, such correlations have also been shown to violate the principle of Information Causality (IC), a generalization of NS principle \cite{Pawlowski2009} (see also \cite{Miklin2021,Naik2022,Patra2023}).

Naturally, the following questions arise: Is it possible to provide an intuitive `explanation' that elucidates the striking features of non-local correlations \cite{Self1}? How does communication complexity become trivial with PR correlation, which is otherwise a NS resource? Why do some non-local correlations violate IC while others are perfectly compatible with it? Interestingly, in this letter, we show that cryptography once again plays a benevolent role in addressing these questions. Inspired by basic ideas in cryptography, we develop a model that provides simple explanations for the aforesaid striking features of non-local correlations. From a foundational standpoint, our model shares similar features with the De Broglie-Bohm theory, as we model non-local boxes as `realistic' systems with instantaneous signaling at the hidden variable level. The superluminal signaling mechanism is designed in such a way that it does not result in signaling at the operational level. Since the design is closely linked to the famous symmetric key encryption model called {\it One-time Pad} (OTP) \cite{Bellovin2011}, we refer to this as the OTP-model for non-local correlations. Before delving into the utility of this model, in the following, we will first formally introduce it and analyze some of its generic aspects.

\section{Preliminaries and set-up}
\subsection{Bipartite no-signaling correlations}
Bipartite no-signalling correlations are joint probability distributions of the form $P_{NS}(a,b|x,y)$ that satisfy the no-signalling (NS) conditions: 
\begin{subequations}
\begin{align}
P_{NS}(a|x,y')&=P_{NS}(a|x,y)\quad \forall~y,y',a;\\
P_{NS}(b|x',y)&=P_{NS}(b|x,y)\quad \forall~x,x',b.
\end{align}  
\end{subequations}
Here, $x\in X$, $y\in Y$, $a\in A$ and $b\in B$ are sample spaces of finite size and $P_{NS}(a|x,y):=\sum_b P_{NS}(a,b|x,y)$ \& $P_{NS}(b|x,y):=\sum_a P_{NS}(a,b|x,y)$ are marginal distributions. These correlations are `Bipartite' as they invoked two parties (say Alice and Bob) that are spatially  separated, and the correlation in this context correspond to the joint probability distribution of outcomes $a$ and $b$ of experiments $x$ and $y$ performed by Alice and Bob, respectively. The no-signalling condition imposes the requirement that in such scenarios, Alice's experimental settings should not influence Bob's measurement statistics and vice-verse.

The classic example of such correlations are the joint probability distribution that violates the Bell's inequality in the standard Bell scenario where local measurements are done by space-like separated parties on a common entangled state. In \cite{Popescu1994}, it was shown that the set of no-signalling correlations is strictly larger than the set of quantum correlations and in particular that there are no-signalling correlations that are more non-local than quantum correlations. A notable example that will be used later on is the PR box, which is a bipartite no-signalling correlations defined as,
\begin{align}
P_{PR}(a,b|x,y)=\frac{1}{2} \delta(a\oplus b,xy);~~a,b,x,y\in \{0,1\}.    
\end{align}
It can be easily verified that this correlation satisfies the NS condition and moreover it achieves the algebraic maximum value of Clauser-Horne-Shimony-Holt (CHSH) expression \cite{Clauser1969}. An important question to ask at this point is whether such correlations can be physical and if not why? An interesting approach in this direction is to look for principles that are physically or information theoretically motivated that we expect to hold. Towards this end, it turns out that the PR box violates an information theoretic principle called the principle of Information Causality (IC) \cite{Pawlowski2009} which will be explained in Section \ref{ICref}. The structure of the set of bipartite no-signalling (NS) correlations is also well known. In particular, the NS conditions being linear constraints on the probability imply that the set of NS correlations form a polytope called NS-polytope. The vertices of the set include deterministic local correlations and correlations that are non-local. For example, in the case of $2$-input, $2$-output correlations (where $a,b,x,y \in \{0,1\}$), all the non-local vertices are equivalent to the PR-box up-to local relabellings.

\subsection{De Broglie-Bohm theory}
Bohm's 1952 pilot-wave model for quantum mechanics \cite{Bohm1952(1),Bohm1952(2)}, which has roots to de Broglie's 1927 work \cite{Broglie1928}, is a hidden variable model of quantum mechanics where each particle is associated with an ontic position and it evolves `classically' under a quantum potential which is related to the amplitude square of the wavefunction. As is famous, this model however allows for instantaneous signalling at the hidden variable level which is washed out by the inherent randomness in the theory (which is also related to the amplitude square of the wavefunction). The curious interplay between this instantaneous signalling and randomness was shown in \cite{Valentini1991}, which proved that if the quantum potential deviates from the amplitude square of the wavefunction, the instantaneous signalling becomes operationally evident. However, the interplay between instantaneous signalling and noise can be more complicated than just one washing out the other as is suggested by the quantum advantages in information processing tasks. However, since the pilate-wave model is originally defined for continuous degrees of freedom, it obscures how signalling and randomness plays together. Moreover, since it is a model for quantum theory, there maybe further obfuscation associated with other features of quantum theory. In this paper, we present a hidden-variable model which is qualitatively similar to the pilate-wave theory but attempts to get around the problems aforementioned. More precisely, we present a model for a class of finite input-output non-local correlations which features the aforementioned interplay between instantaneous signalling and inherent randomness, but is nevertheless less obscured by the fact that it is finite dimensional and has less structure than a model for quantum theory. It goes without saying that this model is less powerful than quantum theory in information processing tasks, but we show that it is precisely the lack of structure than enables us to see the non-trivial ways in which noise and instantaneous signalling inter-plays for non-local advantages in information processing tasks. 

\subsection{OTP-box} 
The idea of cryptography has been instrumental in the development of the current information era \cite{Kahn1996}. OTP is a symmetric encryption scheme where the sender and the receiver share the same key. The key here is a bit string which is long as of the message to be encrypted and is uniformly randomly sampled and, crucially, only used once. The encryption step is simply to `XOR' the message with the key, {\it i.e.}, $\rm{Encrypt} =\rm{Message}\oplus\rm{key}$. The encrypted message is then sent over to the receiver, who will decrypt the message again by `XOR'-ing with the same key, {\it i.e.}, $\rm{Decrypt}=\rm{Encrypt}\oplus\rm{key}=\rm{Message}$. As analyzed by Shannon, the security of this protocol is ensured when the key is sampled uniformly randomly from the set of strings $\{0,1\}^m$, where $m$ is the bit-length of the message \cite{Shannon1949}.
\begin{figure}[t!]
\centering
\includegraphics[scale=0.35]{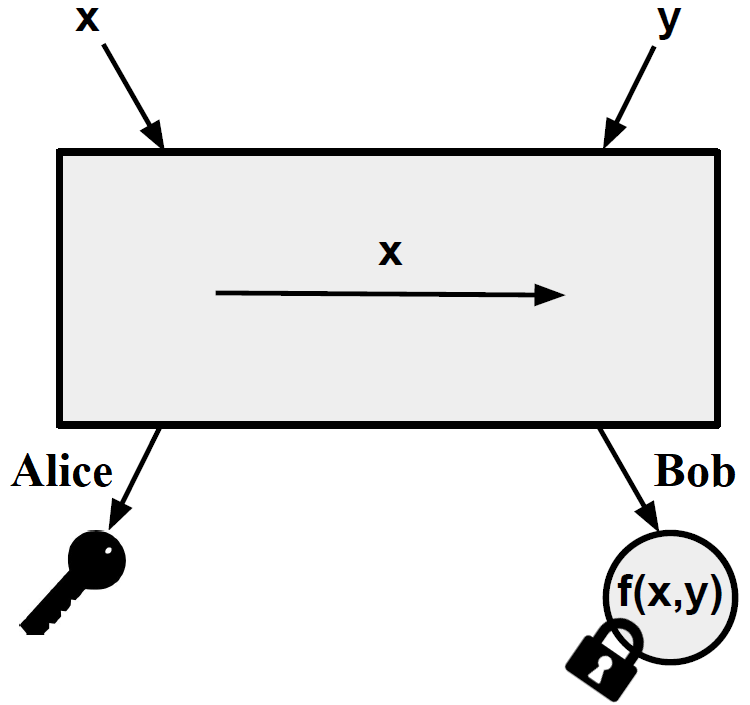}
\caption{An illustration of the OTP box. Given inputs $x$ and $y$ to Alice and Bob, respectively, the box outputs the encrypted form of $f(x,y)$ at Bob's end and the associated {\it key} at Alice's end. In OTP encryption, the {\it key} is drawn uniformly randomly to comply with the operational NS condition.}\vspace{-.5cm}
\label{fig}
\end{figure}

Importantly, as long as the receiver does not have the key, the OTP system is equivalent to a non-signaling scenario in the sense that the receiver gets no information about the message the sender intends to send. This is precisely the observation that prompted our model. In other words, nature can, in principle, perform `secure' instantaneous signaling as long as no one has access to the key a priori. We note that this is analogous to \textit{oblivious embeddings} used in Ref.\cite{obl}. To mathematically capture the connection between NS correlations and OTP-based crypto-system, we start with formally defining an OTP-box. 
\begin{definition}\label{def1}
An OTP-box, shared between two parties (say Alice and Bob), is a black box defined by the tuple $\left\{X,Y,A,B,\Lambda,P,g,f\right\}$, where $X=[m], Y=[n], A=[2]=B$ are sets denoting Alice's and Bob's inputs and outputs, respectively; $g: X \rightarrow A$, $f: X \times Y \rightarrow B$, are functions that define how the inputs are mapped to the outputs; $\lambda\in\Lambda=\{0,1\}$ is the hidden variable that is inaccessible to the agents using the black box and is distributed according to a probability distribution  $\{P(\lambda)\}$. Moreover, the input-output correlation $\mathbf{P}\equiv\{P(a,b|x,y)~|~x\in X, y\in Y,a\in A,b\in B\}$ that the agents observe outside the box is given by, $ 
P(a,b|x,y)=\sum_\lambda P(\lambda)\times\delta(a, g(x)\oplus \lambda)\times\delta(b,f(x,y)\oplus \lambda)$, where $\delta(~,~)$ is the Kronecker delta.
\end{definition}
A quick inspection reveals the inspiration for the definition. In the language of OTP encryption, one may think of $\lambda\in\{0,1\}$ as the {\it key} in the encryption protocol. The OTP box does the following: depending on the input $x$, Alice receives the output $g(x)\oplus \lambda$. Since Alice knows $g(x)$, she effectively has access to the key. Bob's output can be understood as the function $f(x,y)$, but encrypted according to $\lambda$ as $b=f(x,y)\oplus \lambda$ (see Fig.\ref{fig}). Within the black box, the information of Alice's input is instantaneously communicated to Bob's end, albeit inaccessible to Bob outside the box. As we will show now, such crypto-systems satisfy NS property for a particular key distribution. 
\begin{Lemma}\label{lemma1} 
The input-output correlation obtained from an OTP-box satisfies NS if and only if the distribution of the hidden variable within the black box (the key) is uniform, {\it i.e.}, $P(\lambda)=1/2$.
\end{Lemma}
\begin{proof}
The correlation realized by the OTP box is given by,
\begin{align}
P(ab|xy)=\sum_\lambda \delta(a,g(x)\oplus \lambda)\delta(b,f(x,y)\oplus \lambda)P(\lambda)
\end{align}
The form of the output response at the hidden variable level ensures Bob to Alice NS condition of the observed input-output correlation for any distribution $\{p(\lambda)\}_{\lambda=0}^1$ of the hidden variable. To see the NS condition in the other direction, let us consider the Bob's marginal distribution $P_B$:
\begin{align}
&P_B(b|x,y)=\sum_a P(ab|xy)\nonumber\\ 
&=\sum_a \sum_\lambda \delta(a,g(x)\oplus \lambda) \delta(b,f(x,y)\oplus \lambda)P(\lambda) \nonumber\\ 
&=\sum_\lambda \left(\sum_a \delta(a,g(x)\oplus \lambda)\right) \delta(b,f(x,y)\oplus \lambda)P(\lambda) \nonumber\\ 
&=\sum_\lambda \delta(b,f(x,y)\oplus \lambda)P(\lambda).
\end{align}
Thus we have,
\begin{align}
P_B(1|x,y)&=\sum_\lambda (f(x,y)\oplus\lambda)P(\lambda)\nonumber \\
&=f(x,y)\frac{(1+r)}{2}+\overline{f(x,y)}\frac{(1-r)}{2}\nonumber\\
&=\frac{1}{2}+\frac{f(x,y)-\overline{f(x,y)}}{2} r.
\end{align}
Here we use the parameterization $P(\lambda=0)=\frac{1+r}{2}$, with $r\in [-1,1]$. $\overline{z}$ denotes the complement of the bit $z$. Therefore, if $f(x,y)$ explicitly depends on $x$, to ensure NS, we must have $r=0$. This implies that $P(\lambda)=\frac{1}{2}$, and hence proves the Lemma. Note that the result directly follows from Shannon's proof of the unbreakability of OTP encryption scheme \cite{Shannon1949}.
\end{proof}
 
\section{OTP model for nonlocal correlations} 
From now on-wards, we will only consider those OTP-boxes that result in NS correlations. An OTP-box for a NS correlation (violating Bell inequality) can be thought of as a discrete version of the deterministic De Broglie-Bohm model; and interestingly, Lemma \ref{lemma1} here is analogous to the Signal-locality theorem as in the case of original De Broglie-Bohm theory \cite{Valentini1991}. The interplay of signaling and randomness in simulating quantum statistics is also noteworthy at this point \cite{Kar2011,Hall2011,Aravinda2015,obl}.    

For an explicit example we consider arguably the most famous $2,2$-input \& $2,2$-output PR correlation, $P_{\rm{PR}}(a,b|x,y):=\frac{1}{2}\delta(a\oplus b,xy)$, where $a,b,x,y\in\{0,1\}$. An OTP model that reproduces this correlation is given by $g(x)=0$ and $f(x,y)=xy$, and accordingly, Alice's and Bob's outputs from the OTP-box are  $a=\lambda$ and $b=xy\oplus \lambda$ respectively. According to Lemma \ref{lemma1}, $\lambda$ being sampled with uniform probability from the set $\{0,1\}$ ensures NS outside the OTP-box. As pointed out by the author in \cite{Bub2014}, there is a fundamental difference between PR box and its underlying deterministic OTP-box. In a PR box, the temporal order of the inputs is irrelevant, and the correlations arise as a global feature of the statistics. In contrast, the temporal order of the inputs is relevant in OTP-box to determine its output values, which thus requires the assumption of a preferred foliation in space-time. 

OTP model for all the nonlocal vertices of $2,2$-input \& $2,2$-output NS polytope follows immediately as these vertices are nothing but local relabelling of the aforesaid PR correlation \cite{Barrett2005(1)}. In fact, one can consider more generic NS correlations with $m$-input on Alice's side and $n$-input on Bob's side, with each input having binary outputs. A nonlocal vertex of the $m,n$-input \& $2,2$-output NS polytope will be called a full-output vertex if, in the marginal distributions, all possible outputs occur with nonzero probability for all the inputs. Notably, if for one of Alice's inputs, an $m,n$-input NS box only gives one output, one can simulate the box with a $(m-1),n$-input full-output NS box that simply ignores the redundant input. Our next result asserts the OTP model for all the full-output nonlocal vertices of $m,n$-input \& $2,2$-output NS polytope. 
\begin{theorem}\label{theo1}
All the full-output nonlocal vertices of $m,n$-input \& $2,2$-output NS correlations allow an OTP model.
\end{theorem}
\begin{proof}
We start by noting that a complete characterization of nonlocal vertices of $m,n$-input \& $2,2$-output correlations was provided in Ref.\cite{Nick2005}. For our purposes, only two main observations suffice: (i) all the entries of a full-output vertex are either `$0$' or `$1/2$'; (ii) for a full-output vertex, Alice's and Bob's outputs are either strictly correlated or strictly anti-correlated. Let us consider a function $h:X\times Y\rightarrow \{0,1\}$ that indicates whether the outputs will be correlated (`$0$') or anti-correlated (`$1$'), {\it i.e.}, $a\oplus b= h(x,y)$. From Ref.\cite{Barrett2005,Nick2005}, it follows that such a function is unique given a full-output vertex. The correlation corresponding to such a vertex then reads as:
\begin{align}\label{vertex}
P_{\rm vertex}(ab|xy)&=\frac{1}{2}\delta(a\oplus b, h(x,y)),\\ 
&=\sum_\lambda \frac{1}{2}\delta(a,\lambda)~\delta(b,h(x,y)\oplus \lambda).
\end{align}
This is, however, the output correlation of an OTP box with $g(x)=0$, $f(x,y)=h(x,y)$, and $P(\lambda)=\frac{1}{2}$; and thus proves the claim.
\end{proof}

At this point naturally the question arises whether the above theorem can be generalized for higher input-output nonlocal boxes. To this aim we can start with extending the definition for OTP-box appropriately.
\begin{definition}\label{def1a}
A generalised OTP box of type-k (k-GOTP), shared between two parties (say Alice and Bob), is a black box defined by the tuple $\left\{X,Y,A,B,\Lambda,P,g,f\right\}$, where $X=[m], Y=[n], A=[r], B=[s]$ (with $r<s$) are sets denoting Alice's and Bob's inputs and outputs, respectively; $g: X \rightarrow A$, $f: X \times Y \rightarrow B$, are functions that define how the inputs are mapped to the outputs; $\lambda\in\Lambda=[k]$ (with $k\le r$) is the hidden variable that is inaccessible to the agents using the black box and is distributed according to a probability distribution  $\{P(\lambda)\}$. Moreover, the input-output correlation $\mathbf{P}\equiv\{P(a,b|x,y)~|~x\in X, y\in Y,a\in A,b\in B\}$ that the agents observe outside the box is given by, $ 
P(a,b|x,y)=\sum_\lambda P(\lambda)\times\delta(a, g(x)+ \lambda \mod k)\times\delta(b,f(x,y)+ \lambda \mod k)$, where $\delta(~,~)$ is the Kronecker delta.
\end{definition}
Similar interpretation as discussed after Definition \ref{def1} also holds here. For our purposes, it is sufficient to notice that if $f(x,y)$ is a binary function that has a nontrivial dependence on x, then no-signalling condition of a k-GOTP box is equivalent to $P(\lambda)=1/k$. With this, a result analogues to Theorem \ref{theo1} can be stated for the 2,2-input \& m,n-output scenario as well.
\begin{theorem}\label{theorem2}
The nonlocal vertices of $2,2$-input \& $m,n$-output NS correlations allow a k-GOTP model.
\end{theorem}
\begin{proof}
The complete characterization of nonlocal vertices with 2 inputs ({\it i.e.}, $m=n=2$) and r,s-outputs (without any l.o.g. assume $r<s$) is given in Ref.\cite{Barrett2005(1)}. In particular, note that for a given value of $r$ and $s$, there are several inequivalent boxes labelled by $k\in\{2,...,r\}$ as:
\begin{align}
P_k(ab|xy) = 
\begin{cases} 
\frac{1}{k} & : (b - a) \mod k = x \cdot y \\
& \quad a, b \in \{0, \ldots, k-1\} \\
0 & : \text{otherwise}.
\end{cases}
\end{align}
Clearly each of these boxes can be modelled as a k-GOTP box with $f(x,y)=x.y$ and $g(x)=0$. Thus we have,
\begin{align}
P_k(ab|xy)=\frac{1}{k}\sum_\lambda \delta(a,\lambda)\times \delta(b,x.y+\lambda \mod k).
\end{align}
This completes the proof.
\end{proof}
At this point, we recall that quantum correlations lie within the NS polytope. Therefore, to include quantum realizable and physically interesting NS correlations within OTP model framework, we proceed to introduce a generalized notion of the OTP-box.
\begin{definition}
A noisy OTP-box (N-OTP-box) is defined by the tuple $\left\{X,Y,A,B,\Lambda_1,\Lambda_2,P,g,f\right\}$, where $X=[m], Y=[m], A=[2]=B$ are sets denoting Alice's and Bob's input and output sets respectively; $g: X \rightarrow A$, $f: X \times Y \rightarrow B$ are functions that define how the inputs are mapped to the outputs; $\lambda_1\in\Lambda_1=\{0,1\}=\Lambda_2\ni\lambda_2$ are hidden variables that are inaccessible to the agents using the black box and is distributed according to $P(\lambda_1,\lambda_2)$. Moreover, the input-output correlation $\mathbf{P}$ observed by the agents outside the N-OTP-box is given by, $P(a,b|x,y)=\sum_{\lambda_1,\lambda_2}P(\lambda_1,\lambda_2)\times \delta(a,g(x)\oplus \lambda_1)~\times\delta(b,f(x,y)\oplus \lambda_2)$.
\end{definition}
A generalization of Lemma \ref{lemma1} ensures the input-output correlation obtained from an N-OTP-box to be NS whenever $P(\lambda_2)=\sum_{\lambda_1}P(\lambda_1,\lambda_2)=1/2$. Note that N-OTP-boxes reduce to the corresponding OTP-boxes when $P(\lambda_1,\lambda_2)=\frac{1}{2}\delta(\lambda_1,\lambda_2)$, {\it i.e.}, the variables $\lambda_1$ and $\lambda_2$ are perfectly correlated. In the absence of perfect correlation between $\lambda_1$ and $\lambda_2$, an N-OTP-box can be interpreted as follows: output at Bob's end is $f(x,y)$ but encrypted by the key $\lambda_2$, i.e. $f(x,y)\oplus \lambda_2$. However, unlike the perfect case, the output at Alice's end ($\lambda_1$) is not exactly the key to Bob's encrypted output but correlated to it. Consequently, Bob cannot perfectly decrypt the value of $f(x,y)$. As shown in the following result this noisy box leads to the construction of the OTP model for non-extremal NS correlations. 
\begin{theorem}\label{theo2}
All isotropic $2,2$-input \& $2,2$-output NS correlations allow an N-OTP model.   
\end{theorem}
\begin{proof}
The isotropic correlations are of the form, $P_{\rm{ISO}}:=q P_{\rm{PR}}+(1-q) P_{\rm{anti-PR}}$, where $P_{\rm{anti-PR}}(a,b|x,y):=\frac{1}{2}\delta(a\oplus b,xy\oplus1)$. The fact that both PR and anti-PR correlations have OTP models implies:
\begin{align*}
P_{\rm{ISO}}&=q P_{\rm{PR}}+(1-q) P_{\rm{anti-PR}}\\
&=\frac{1}{2}\sum_\lambda[q \delta(a,\lambda)+(1-q) \delta(a,\lambda\oplus 1)]\times\delta(b,xy\oplus\lambda)\\
&=\sum_{\lambda_1,\lambda_2}p(\lambda_1,\lambda_2)\times\delta(a,\lambda_1)\times\delta(b,xy\oplus \lambda_2), 
\end{align*}
where, $p(\lambda_1,\lambda_2)=\frac{1}{2}[q\delta(\lambda_1,\lambda_2)+(1-q)\delta(\lambda_1\oplus1,\lambda_2)]$. The last expression is in the N-OTP form, and hence the claim is proved.
\end{proof}

\section{OTP Model: Implications}
As the basic tenets of the OTP model framework are developed and the model is generalized to some extent, we now move on to showcase the benefits of this model in explaining and understanding some of the striking features of non-local correlations.
\subsection{Collapse of communication complexity} 
Nonlocal correlations can significantly reduce the communication complexity of distributed computing \cite{Buhrman2010}. Given an $m$-bit string $x$ to Alice and an $n$-bit string $y$ to Bob, any arbitrary binary function $f(x,y):\{0,1\}^m\times\{0,1\}^n\mapsto\{0,1\}$ can be computed by Bob with just $1$-bit of communication from Alice to Bob provided they have access to nonlocal correlations. However, the completion of a task demanding m-bit communication using only a 1-bit channel, aided by nonlocal correlations that are otherwise NS, lacks a straightforward, intuitive explanation.

The OTP model framework offers an innate explanation for this puzzle. Let as a resource Alice and Bob are given an OTP-box where $g(x)=0$ and $f(x,y)$ in Def.\ref{def1} is precisely the function they are interested in computing. From Lemma \ref{lemma1}, this simply means that they are given the appropriate NS correlation. Alice and Bob input $x$ and $y$ in their respective parts of the box. According to the symmetric encryption scheme of the OTP-box, Bob's output is the encrypted version of the function $f(x,y)$, {\it i.e.}, $b=f(x,y)\oplus \lambda$, whereas Alice output is the key $a=\lambda$. This solves the puzzle as to how Alice and Bob can compute a function that requires communication of $m$-bit with just $1$-bit of communication and a NS box. The information of $x$ reaches Bob's end through superluminal signaling at the hidden variable level. The output of the box is, however, encrypted and hence useless to Bob. But Alice has the key. She now sends the key ($\lambda$) through a $1$-bit communication channel, which allows Bob to decrypt the value of the function $f(x,y)$, irrespective of the length of the string $x$. In summary, we can say that the key is used by the agents to {\it unlock} the power of superluminal signaling at the hidden variable (ontic) level.

\subsection{Van Dam protocol} 
Win van Dam has shown that PR correlation can trivialize communication complexity \cite{vanDam2005}. This is quite surprising when compared with quantum correlations. Quantum entanglement is known to reduce the amount of classical information that Alice and Bob need to exchange to evaluate certain distributed functions \cite{Cleve1997}, whereas for other functions entanglement provides no advantage. An example of latter is the Inner Product function $\mathrm{IP}_n:\{0,1\}^n\times\{0,1\}^n\mapsto\{0,1\}$, which is defined as $\mathrm{IP}_n(x,y):=\oplus_{i=1}^nx_i\cdot y_i$ \cite{Cleve1998}. At this point we recall a cryptographic concept called homomorphic encryption \cite{Fontaine2007}. Homomorphic encryption allows performing computations on encrypted data without decrypting it and hence without compromising the security of the data. Our next result provides a new outlook of Van Dam protocol from a homomorphic encryption point of view. 
\vspace{-.15cm}
\begin{proposition}\label{prop1}
The Van Dam protocol for distributed computing of arbitrary binary functions with trivial communication complexity can be understood as an application of the homomorphic nature of OTP encryption with respect to XOR operation.
\end{proposition}
\begin{proof}
The simplest example of homomorphic encryption is the OTP encryption, which is homomorphic with respect to modular addition, {\it i.e.}, it is additively homomorphic. For instance, consider the "(encrypted data, key)" pairs $(e=z\oplus \lambda,\lambda)$ and $(e^\prime=z^\prime\oplus \lambda^\prime,\lambda^\prime)$. XOR-ing operation on the encrypted data and the key yields: $(e\oplus e^\prime,\lambda\oplus\lambda^\prime)=(e\oplus e^\prime= z\oplus \lambda\oplus z^\prime\oplus\lambda^\prime, \lambda\oplus\lambda^\prime)=(e\oplus e^\prime= (z\oplus z^\prime)\oplus(\lambda\oplus\lambda^\prime), \lambda\oplus \lambda^\prime)$. Thus, post-computation, the new key ($\lambda+\lambda^\prime$) decrypts the new encrypted data ($(z\oplus z^\prime)\oplus(\lambda\oplus\lambda^\prime)$). In other words, it is possible to perform modular addition of the data without decrypting it first.   

Coming to think of it, this is precisely what happens in Van Dam's protocol. In the distributed computing scenario Van Dam considered, Alice is given some string $x\in X\equiv\{0,1\}^m$ and Bob is given another string $y\in Y\equiv \{0,1\}^n$. Their task is to compute some arbitrary binary function $f(x,y)$ that is apriori known to the players. Let us now express Van Dam's protocol in the OTP-box model. For every possible $y_i \in Y\equiv\{0,1\}^n$, Alice inputs $f(x,y_i)$ into a PR box, and thus requiring a total of $|Y|=2^n$ PR boxes. Bob, on the other hand, inputs `1' into the $y_{\rm{dec}}^{th}$ PR box, where $y_{\rm{dec}}$ is the decimal representation of the binary string `$y$' (or example, if $y=101$, then Bob inputs `1' into $y_{dec}=5$th box). Bob inputs `$0$' into all the remaining PR boxes. Modeled as an OTP box, the outputs at Bob's end are the encrypted information $f(x,y_i)\delta(i,y_{\rm{dec}})\oplus \lambda_i$. Here, $\delta(i,y_{\rm{dec}})$ is Bob's input into the $i^{th}$ PR box. Now, note that $\oplus_i f(x,y_i)\delta(i,y_{\rm{dec}}) =f(x,y)$. Thus, if Bob can decrypt the output of each PR box, Bob can compute $f(x,y)$. But this requires  Alice to send each $\lambda_i$ to Bob. Fortunately, the additive-homomorphism of OTP encryption comes to the rescue! Bob knows that OTP encryption is homomorphic with respect to XOR-ing operation. He thus blindly XORs all the encrypted bits to obtain the bit $\oplus_i (f(x,y_i)\delta(i,y_{\rm{dec}})\oplus \lambda_i) = (\oplus_i f(x,y_i)\delta(i,y_{\rm{dec}})) \oplus (\oplus_i \lambda_i)=f(x,y) \oplus (\oplus_i \lambda_i)$. Meanwhile, Alice computes the new key $\lambda=\oplus_i \lambda_i$ and sends this bit to Bob. The homomorphism then allows Bob to decrypt the value of the function $f(x,y)$.
\end{proof}
Casting Van Dam's protocol as homomorphic encryption thus provides a better understating of distributed computations with nonlocal correlations. Accordingly, the puzzling feature of the breakdown of communication complexity with PR box can be better appreciated in the OTP framework. 

\subsection{The principle of Information Causality} \label{ICref}
While nonlocality in PR correlation is quite strong, making communication complexity trivial, nonlocality in quantum theory is docile enough to ensure `nontrivial communication complexity' \cite{Brassard2005}. Lately, a novel principle called Information Causality (IC) was proposed, which can capture this limited nonlocality of quantum theory and can identify post-quantum nonlocal correlations. IC limits Bob’s information gain about a previously unknown to him dataset of Alice to be at most $m$ bits when Alice communicates $m$ classical bits to Bob, and he is allowed to use local resources that might be correlated with Alice \cite{Pawlowski2009}. A necessary condition of IC can be obtained by considering random access code (RAC) task \cite{Wiesner1983,Ambainis1999,Ambainis2002}, where Alice receives a random string $x\in\{0,1\}^n$, and Bob randomly aims to guess $y^{th}$ bit $x_y$ of Alice's string. In the presence of an NS correlation, if Alice communicates $m$-cbits to Bob, then according to IC the efficiency of the aforesaid task is upper bounded by $I_n\le m$, where $I_n:=\sum_{k=1}^nI(x_k=\beta|y=k)$. Here, $I(x_k=\beta|y=k)$ denotes Shannon's mutual information between the $k^{th}$ bit of Alice, and Bob’s guess $\beta$. Interestingly, IC reproduces the Cirel'son bound \cite{Cirelson1980}, {\it i.e.}, any nonlocal correlation yielding CHSH \cite{Clauser1969} value more than $2\sqrt{2}$ violates IC.

A natural question is: how is this limited nonlocal feature of quantum correlation  captured in our OTP model framework? Nonlocality, in our model, is achieved by a careful interplay between ontic-level signaling and randomness. Consequently, limited nonlocality can be accounted for in two ways: (i) either messing around with the randomness, which boils down to noisy encryption, or (ii) reducing the amount of ontic-level signaling. We will examine both these approaches in reproducing quantum correlations, particularly focusing on the class of isotropic correlations.

{\it (i) Quantum correlations as noisy encryption:} As shown in Theorem \ref{theo2}, isotropic correlations in $2,2$-input \& $2,2$-output scenario allow an N-OTP model. As motivated in the definition of the N-OTP box, the limited non-locality found in nature can be explained through noisy encryption. More precisely, we can interpret it as follows: there is a well-defined value of the encryption key $\lambda$ within the box, but nature is `peculiar' in such a way that it does not provide perfect access to the key to the agent outside the box (in this case Alice). The key $\lambda^\prime$ accessed by Alice is not identical to the key $\lambda$ within the box; rather, they are correlated. A simple calculation ensures that the necessary condition of IC is satisfied when the mutual information between the true key $\lambda$ and the obtained key $\lambda^\prime$ is upper bounded by $0.5$ bit. Although this could be a way of obtaining quantum-realizable nonlocal correlation, the approach might look rather arbitrary and {\it ad hoc}.

{\it (ii)  Quantum correlations through limited ontic-level signalling:} A more natural approach to explain limited nonlocality would be to limit the extent of instantaneous signaling at the ontic level. After all, the idea of instantaneous signaling itself sounds far too radical. Let us try to come to terms with it by making it less powerful. For this, we may model the instantaneous communication within the box as a classical binary bit-flip channel, with flipping probability $(1-\mu)$. The remainder of the OTP-box remains the same. In this case, the input-output correlation obtained by the agents outside the box reads as:
$P(ab|xy)=\mu \frac{1}{2}[\sum_\lambda \delta(a,\lambda)\delta(b,xy\oplus\lambda)]+(1-\mu) \frac{1}{2}[\sum_\lambda \delta(a,\lambda)\delta(b,(x\oplus 1)y\oplus\lambda)]$. We can now use the standard RAC protocol to verify that OTP boxes of this kind always violate the IC principle, except when $\mu=0.5$, corresponding to no communication at all. In other words, perfect encryption, with noisy ontic-level signaling, violates IC. By perfect encryption, we mean that Alice has perfect access to the key that is used to encrypt the function $f(x,y)$ at Bob's end. A more explanatory analysis is provided in Appendix.

\section{Discussions} 
In summary, we have constructed a hidden variable model for nonlocal correlations. From a foundational point of view, our model is analogous to the De Broglie-Bohm model that allows instantaneous signaling at the hidden variable/ontic level. Interestingly, the model displays an elegant interplay between ontic-level signaling and randomness, {\it i.e.}, distribution of the ontic variable. Randomness essentially solves the tension between ontic-level signaling and the operational NS principle by `toning down' the power of ontic-level signaling at the operational level. From a cryptographic point of view, our model can also be seen as a symmetric key encryption system called One Time Pads. This, in turn, helps us to explain the breakdown of communication complexity with PR correlation and provides a better understanding of Van Dam's protocol in terms of homomorphic encryption. As it turns out, the ontic variables in our model play the role of the encryption key, and the randomness in its distribution is designed in such a way that one can, nevertheless, {\it unlock} some power of the ontic-level signaling, leading to an advantage in nonlocal computation. In addition to its foundational implication, this shows an instance of the utility of the connection we presented between encryption and nonlocal boxes. We further believe that the results in classical cryptography can be used to greatly improve our understanding of nonlocal correlations and also in designing new protocols that can exploit this nonlocality.
We also discussed possible ways of obtaining the limited nonlocal behavior of quantum correlations within this cryptographic ontic model framework. As it turns out, by making the ontic-level signaling noisy, one cannot recover quantum correlations. This suggests that within the OTP framework, noisy encryption is necessary for quantum realisability. That is, the randomness at Alice's and Bob's end cannot be highly correlated. This is an added constraint imposed on the randomness by IC, in addition to the `uniform marginals' constraint imposed by NS.

Apart from trivializing communication complexity, post-quantum nonlocal correlations are also known to exhibit some other exotic features. For instance, in the zero-error communication scenario where the aim is to transmit messages through this channel with no probability of confusion \cite{Shannon1956,Cubitt2010,Alimuddin2023}, there exist NS correlations that can yield positive zero-error capacity even for channels with no unassisted capacity \cite{Cubitt2011}. An OTP model explanation might provide a better understanding of this phenomenon. Another esoteric problem where we expect this model to provide insights into is nonlocality distillation \cite{Forster2009,Eftaxias2023,Naik2023,Ghosh2024}. We believe that the existing literature on cryptography over noisy channels can provide insights into nonlocality distillation when one thinks of nonlocal quantum correlations as OTP-boxes with noisy encryption. Finally, we note that this connection between nonlocality and cryptography is not only limited to OTP encryption, and therefore, qualitatively different encryption schemes may also add to our understanding of nonlocality.

\begin{acknowledgments}
We are thankful to Nicolas Gisin, S Aravinda, and Jebarathinam Chellasamy for their suggestions on the earlier version of the manuscript. We gratefully acknowledge fruitful discussions with Mir Alimuddin, Ananya Chakraborty, Snehasish Roy Chowdhury, Guruprasad Kar, Amit Mukherjee, Sahil Gopalkrishna Naik, Ram Krishna Patra, and Samrat Sen. MB acknowledges funding from the National Mission in Interdisciplinary Cyber-Physical systems from the Department of Science and Technology through the I-HUB Quantum Technology Foundation (Grant No. I-HUB/PDF/2021-22/008).%, support through the research grant of INSPIRE Faculty fellowship from the Department of Science and Technology, Government of India, and the start-up research grant from SERB, Department of Science and Technology (Grant No. SRG/2021/000267). 
\end{acknowledgments}

\appendix
\section{IC and N-OTP model}
\vspace{-.25cm}
{\it (i) Quantum Correlations from noisy encryption:} Isotropic correlations, expressed as an N-OTP box, has the form:
\begin{align}\label{Iso}
P(ab|xy)=\sum_{\lambda_1,\lambda_2} \delta(a,\lambda_1)~\delta(b,xy\oplus \lambda_2)~P(\lambda_1,\lambda_2),
\end{align}
where $P(\lambda_1,\lambda_2)$ has uniform marginals, and consequently we have
\begin{align}
P(\lambda_1,\lambda_2)&=\frac{1}{2}P(\lambda_2|\lambda_1)\nonumber\\
&=\frac{\mu}{2}\delta(\lambda_1,\lambda_2)+\frac{1-\mu}{2}\delta(\lambda_1,\overline{\lambda_2}),    
\end{align}
with $\mu\in[0,1]$. Consider the case where Alice and Bob are playing the $2\rightarrow1$ RAC game, where Bob wishes to perfectly guess one of the two bits of the string $x_0x_1$ given to Alice, with the assistance of 1 c-bit of communication. Moreover, let us use the correlation of Eq.(\ref{Iso}) as an assistance to the 1 $c$-bit communication line from Alice to Bob. Following the standard RAC protocol, Alice inputs $x_0\oplus x_1$ into her end of the box. The output at Alice's end is $a=\lambda_1$ and she communicates 
\begin{align}
a\oplus x_0=\lambda_1\oplus x_0  
\end{align}
over to Bob. Bob inputs $y\in\{0,1\}$ into his end of the box if he want's to know the $(y+1)^{\rm th}$ bit of Alice's string. Bob receives the output 
\begin{align}
b=(x_0\oplus x_1)y\oplus \lambda_2    
\end{align}
from the box and accordingly guesses 
\begin{align}
z&=b\oplus (x_0\oplus \lambda_1)\nonumber\\
&=(x_0\oplus x_1)y\oplus\lambda_2\oplus x_0\oplus \lambda_1\nonumber\\
&=x_0\overline{y}\oplus x_1y\oplus \lambda_1\oplus\lambda_2.  
\end{align}
It is clear that whenever $\lambda_1$ and $\lambda_2$ are perfectly correlated, Bob can perfectly guess either of the bits in Alice's string, which implies a violation of the IC principle. Let us study the joint distribution of Bob's output ($z$) and Alice's first bit $x_0$, when Bob wants to predict the first bit (corresponding to $y=0$). This joint distribution is given by,
\begin{align}
&P(x_0,z|y=0)=\sum P(x_0,x_1,z,\lambda_1,\lambda_2|y=0)\nonumber\\
&=\sum P(z|x_0,x_1,\lambda_1,\lambda_2,y=0)P(x_0,x_1,\lambda_1,\lambda_2|y=0)\nonumber\\
&=\sum\delta(z,x_0\oplus\lambda_1\oplus\lambda_2)P(x_0,\lambda_1,\lambda_2|y=0)\nonumber\\
&=\frac{1}{2}\sum \delta(z,x_0\oplus\lambda_1\oplus\lambda_2)P(\lambda_1,\lambda_2)\nonumber\\
&=\frac{\mu}{2}\delta(z,x_0)+\frac{1-\mu}{2}\delta(z,\overline{x_0}).
\end{align}
Here, we used the fact that when $y=0$, then $z=x_0\oplus\lambda_1\oplus\lambda_2$ as discussed earlier. The independence of $x_0$, $\lambda_i$ and $y$ with each other allows us to write $P(x_0,\lambda_1,\lambda_2|y=0)=P(x_0|\lambda_1,\lambda_2,y=0)P(\lambda_1,\lambda_2|y=0)=P(x_0)p(\lambda_1,\lambda_2)=\frac{1}{2}p(\lambda_1,\lambda_2)$. Also note the implicit summation over $x_1$ used in removing the $x_1$ variable from the joint distribution in the third step. The Shannon entropy now takes the form:
\begin{align}
&H(X_0,Z|Y=0)\nonumber\\
&=-\sum_{x_0,z} P(x_0,z|y=0)\log(P(x_0,z|y=0))\nonumber\\
&=1+h(\mu),
\end{align}
where, $h(\mu):=-\mu\log\mu-(1-\mu)\log(1-\mu)$. It is immediate that $H(X_0|Y=0)=H(Z|Y=0)=1$. Therefore, the mutual information is given as:
\begin{align*}
I(X_0:Z|Y=0)&=H(X_0|Y=0)+H(Z|Y=0)\\
&~~~~~~-H(X_0,Z|Y=0)=1-h(\mu).
\end{align*}
Same is the case when $y=1$, and therefore, the sum of the conditional entropies ($I_2$), which is the subject if IC principle, takes the form $I_2=2-2h(\mu)$. Now, IC is violated if $I_2=2-2h(\mu)>1$. Thus, IC is violated if $h(\mu)<\frac{1}{2}$, corresponding to high correlation between $\lambda_1$ and $\lambda_2$. Moreover, one can immediately show that $I(\lambda_1:\lambda_2)=1-h(\mu)$. Thus, in terms of mutual information between the two variables, IC will be violated if $I(\lambda_1:\lambda_2)>1/2$.

{\it (ii) Quantum correlations from noisy ontic-signalling:} Here, the output correlations is given by
\begin{align*}
P(ab|xy)&=\frac{\mu}{2}\sum_\lambda\delta(a,\lambda)\delta(b,xy\oplus\lambda)\\
&~~~~~~~~~~~+\frac{1-\mu}{2}\sum_\lambda\delta(a,\lambda)\delta(b,\overline{x}y\oplus\lambda).
\end{align*}
Unlike earlier, the keys here are perfectly correlated, but the hidden variable communication is noisy. Following the same protocol as before, one can see that Bob's output is given as $z=cy\oplus x_0$. Thus, it is immediate that whenever $y=0$, the output $z$ is fully correlated with $x_0$. That is $I(x_0:z|y=0)=1$. Now, when $y=1$, the answer $z$ equals $x_1$ with probability $\mu$ and it will be wrong with probability $1-\mu$. That is, $I(x_1:z|y=1)=1-h(\mu)$. Thus, we have $I_2=2-h(\mu)>=1$ always violating IC principle, unless $\mu=0.5$.

\end{document}